\documentclass[aps, prb,a4paper,showpacs,reprint]{revtex4-1}
\usepackage{graphicx}
\usepackage{color}
\usepackage{comment}
\usepackage{array}
\usepackage{amsmath}
\usepackage{upgreek}
\usepackage{color,soul}

\definecolor{gray}{gray}{0.6}
\setulcolor{yellow}

\newcolumntype{C}[1]{>{\centering\let\newline\\\arraybackslash\hspace{0pt}}p{#1}}

\begin{document}

\title{Thermal boundary resistance at Si/Ge interfaces determined by approach-to-equilibrium molecular dynamics simulations}
\author{Konstanze R. Hahn}
\email[e-mail: ]{konstanze.hahn@dsf.unica.it}
\author{Marcello Puligheddu}
\author{Luciano Colombo}
\affiliation{Dipartimento di Fisica, Universit\`a di Cagliari\\ Cittadella Universitaria, I-09042 Monserrato (Ca), Italy}

%\pacs{65.40.-b, 63.22.Np}
\begin{abstract}
The thermal boundary resistance of Si/Ge interfaces as been determined using approach-to-equilibrium molecular dynamics simulations. Assuming a reciprocal linear dependence of the thermal boundary resistance, a length-independent bulk thermal boundary resistance could be extracted from the calculation resulting in a value of 3.76$\times 10^{-9}$ m$^2$K/W for a sharp Si/Ge interface and thermal transport from Si to Ge. Introducing an interface with finite thickness of 0.5 nm consisting of a SiGe alloy, the bulk thermal resistance slightly decreases compared to the sharp Si/Ge interface. Further growth of the boundary leads to an increase in the bulk thermal boundary resistance. When the heat flow is inverted (Ge to Si), the thermal boundary resistance is found to be higher. From the differences in the thermal boundary resistance for different heat flow direction, the rectification factor of the Si/Ge has been determined and is found to significantly decrease when the sharp interface is moderated by introduction of a SiGe alloy in the boundary layer.
\end{abstract}

\maketitle

\section{Introduction}
Thermoelectricity as alternative energy production has gained increased interest in recent years and substantial research has emerged to increase the figure of merit $ZT$ of thermoelectric materials.\cite{Dresselhaus2007,Minnich2009,Karni2011}
%The achievable efficiency of latest thermoelectric devices, however, is still very poor. 
The figure of merit describes the efficiency of a thermoelectric material and depends on its Seebeck coefficient and its electric and thermal conductivity. A common approach to achieve a high $ZT$ is by minimization of the thermal conductivity of semiconducting materials and preserving good detailed conduction properties.\cite{Dresselhaus2007,Snyder2008,Minnich2009,Yu2010}

Introduction of impurities to pristine bulk materials by alloying has been shown to be a promising way to minimize the material thermal conductivity $\kappa$. In fact, the thermal conductivity of Si$_x$Ge$_{1-x}$ alloys is reduced remarkably compared to their bulk counterparts already at very small impurity concentrations $x$.\cite{Stohr1939,Abeles1963,Wang2008a,Garg2011,Melis2014b} Similar effects have been shown in Bi$_2$Te$_3$, Bi$_x$Sb$_{2-x}$Te$_3$ and PbTe.\cite{Minnich2009,Dresselhaus2007} Scattering of short-wavelength phonons on impurity atoms is responsible for the decrease of $\kappa$ while thermal transport by mid- and long-wavelength phonons remains unperturbed by atomistic defects.\cite{Minnich2009}
It is therefore essential to suppress the propagation of phonons with longer wavelength for achieving a further reduction of $\kappa$.
%Thus, further minimization of $\kappa$ can be obtained by suppression of the propagation of phonons with longer wavelength.
A possible way to realize this is by nanostructuring of the material confirmed by several experimental and theoretical studies.

%Molecular dynamics simulations have shown, for example, a drastic suppression of low-frequency (long-wavelength) phonons when nanoscopic holes are introduced in a SiGe bulk alloy.
Recently, introduction of nanoscopic holes in a SiGe bulk alloy simulated by molecular dynamics has been shown to drastically decrease the number of low-frequency (long-wavelength) phonons resulting in a remarkable reduction of the thermal conductivity in the SiGe alloy.\cite{He2011}
Furthermore, a lower thermal conductivity in Si/Ge-based materials was observed in either superlattices or nanowires and nanodots depending on the dimension of the nanostructures.\cite{Savic2013}
Similar results have been shown in experimental studies where the thermal conductivity of bulk Si and bulk SiGe alloys was reduced significantly when nanocrystalline structures have been generated by ball milling.\cite{Bathula2012,Wang2008a,Bux2009}
The decrease of $\kappa$ in such nanostructured materials results from increased phonon scattering at the interfaces introduced by nanograins, nanowires or superperiodicity.

The effect of interfaces to thermal conductivity is denoted as thermal boundary resistance (TBR), also known as Kapitza resistance.\cite{Kapitza1941} Yet, a complete understanding of the scattering properties of the interface affecting the TBR and an accurate prediction of such is still a matter of investigation.
Several models have been proposed for the determination of TBR such as the diffuse and the acoustic mismatch model (DMM and AMM, respectively).\cite{Little1959,Swartz1989} However, within these models assumptions of the scattering processes are made a priori. Phonon interface scattering, for example, is assumed to be elastic in both methods. Furthermore, the DMM considers only diffuse scattering while in the AMM diffuse scattering is neglected and only specular scattering is assumed. The TBR values obtained from these models are in general lower than in experiments; for instance, in the example of Si/SiO$_2$ interfaces experimental values range from 2$\times10^{-9}$ to 2$\times10^{-8}$ m$^2$K/W\cite{Lee1997,Kato2008,Hurley2011} while theory predicts a TBR of 2.4-3.5 $\times 10^{-9}$ m$^2$K/W.\cite{Hu2001}

Another possible way to predict TBR is based on molecular dynamics (MD) simulations.
Determination of the TBR of Si/Ge interfaces using non-equilibrium MD simulations resulted in a value of 1.26 $\times 10^{-9}$ m$^2$K/W at a process temperature of 300 K.\cite{Balasubramanian2011} In this study, the thermal boundary resistance has been determined at a finite simulation cell size of 20 nm which can significantly affect the thermal transport properties as has been shown for the length dependent thermal conductivity of bulk materials.\cite{Schelling2002,Sellan2010,Melis2014a} A higher value (2.72-3.17 $\times 10^{-9}$ m$^2$K/W) has been reported by Landry et al. which was reasonably close to the TBR predicted by DMM (2.4 $\times 10^{-9}$ m$^2$K/W).\cite{Landry2009a} Moreover, non-equilibrium MD simulations have recently been used to determine the TBR in SiGe nanowires.\cite{Rurali2014}

The TBR can be affected by the direction of the heat flux as a result of the asymmetry in thermal conductivity of the bulk materials Si and Ge which is defined in the rectification factor.\cite{Dames2009} Depending on the mass and lattice mismatch, the rectification of a two segment bar can change significantly.\cite{Dames2009} Therefore, the quantification of rectification is also of great interest in the characterization of the TBR of a certain material pair.

Here, we have used approach-to-equilibrium MD (AEMD) simulations to estimate the TBR of Si/Ge interfaces. First, we have derived an expression for the length-independent TBR at the theoretical limit of an infinite sample length (section \ref{sec:tbr_theory}). Based on this expression, the TBR of a sharp Si/Ge interface has been evaluated for heat flux both from Si to Ge and vice versa (section \ref{sec:tbr_sharp}). Furthermore, the effect of boundary thickness has been investigated for a boundary thickness up to 2 nm (section \ref{sec:d_effect}). Finally, the rectification of the interface has been determined as a function of the boundary thickness by comparison of the TBR for heat flux from Si to Ge and from Ge to Si (section \ref{sec:rect}).

\section{Methods}
\label{sec:methods}
\subsection{AEMD simulations}

Molecular dynamics simulations have been performed using the \textsc{lammps} code.\cite{Plimpton1995,Plimptop} Interatomic forces have been described applying the Tersoff pair potential\cite{Tersoff1989} which has been shown to represent reasonably well the mechanical and thermal properties of Si and Ge materials.\cite{Melis2014b,Cook1993,Porter1997,Li2012,Li2012a}

The overall thermal conductivity of the simulated systems has been determined using the AEMD method.
Originally, the AEMD method was developed under the assumption that $\kappa$ is uniform in the direction of the heat transport. This is obviously not the case for the here investigated systems. However, from considerations of energy conservation, we can argue that heat transport between two homogeneous systems, where $\kappa$ is stepwise uniform, can indeed be described with an overall thermal conductance coefficient $h_c$ according to

\begin{equation}\label{eq:hc}
\begin{split}
\dot{q} &=h_c \cdot \Delta T\\
\text{with } h_c &=\frac{L_1}{\kappa_1}+R+\frac{L_2}{\kappa_2}
\end{split}
\end{equation}

Assuming such an overall, uniform thermal conductance coefficient for the differential heat transport, the same procedure and equations of the AEMD approach for homogeneous systems can be applied for heat transport through interfaces with stepwise uniform thermal conductivities, supporting the use of the latter also for heterogeneous systems where an overall constant thermal conductivity coefficient $h_c$ can be applied. This assumption has been used previously for the calculation of the thermal conductivity and boundary resistance of Si/SiO$_2$ interfaces.\cite{Lampin2012} Furthermore, we anticipate a result extensively described in section \ref{sec:tbr_sharp}, namely: for a sharp interface, AEMD results are in excellent agreement with other theoretical predictions.

Using the AEMD method, the simulation cells are firstly divided into two regions with comparable length in the direction of thermal transport. One of these two compartments is equilibrated at high temperature ($T_\mathrm{h}$=400 K), the other compartment at low temperature ($T_\mathrm{c}$=200 K) using velocity rescaling. This creates an initial step-like function of the temperature along the sample length in $z$-direction.\cite{Melis2014a} Next, the evolution of the average temperature in the hot ($T_\mathrm{h}$) and cold ($T_\mathrm{c}$) reservoir has been recorded during a transient regime towards equilibrium of microcanonical evolution. Based on Fourier's theorem of thermal transport and the given step-like initial temperature profile, the evolution of the temperature gradient ($\Delta T = T_\mathrm{h}-T_\mathrm{c}$) follows 
\begin{equation}\label{eq:t-grad}
\Delta T\left( t \right)=\sum_{n=1}^\infty C_n e^{-\alpha_n^2\bar{\kappa} t},
\end{equation}
where $\bar{\kappa}=\frac{\kappa}{\rho c_v}$ is the thermal diffusivity with the density $\rho$ of the material and its specific heat $c_v$.
This expression is fitted to the temperature gradient obtained from the simulations to determine $\kappa$. More details on the methodology can be found elsewhere.\cite{Melis2014b,Lampin2013,Melis2014a}

\begin{figure}[tb]
\centering
  \includegraphics[width=0.45\textwidth]{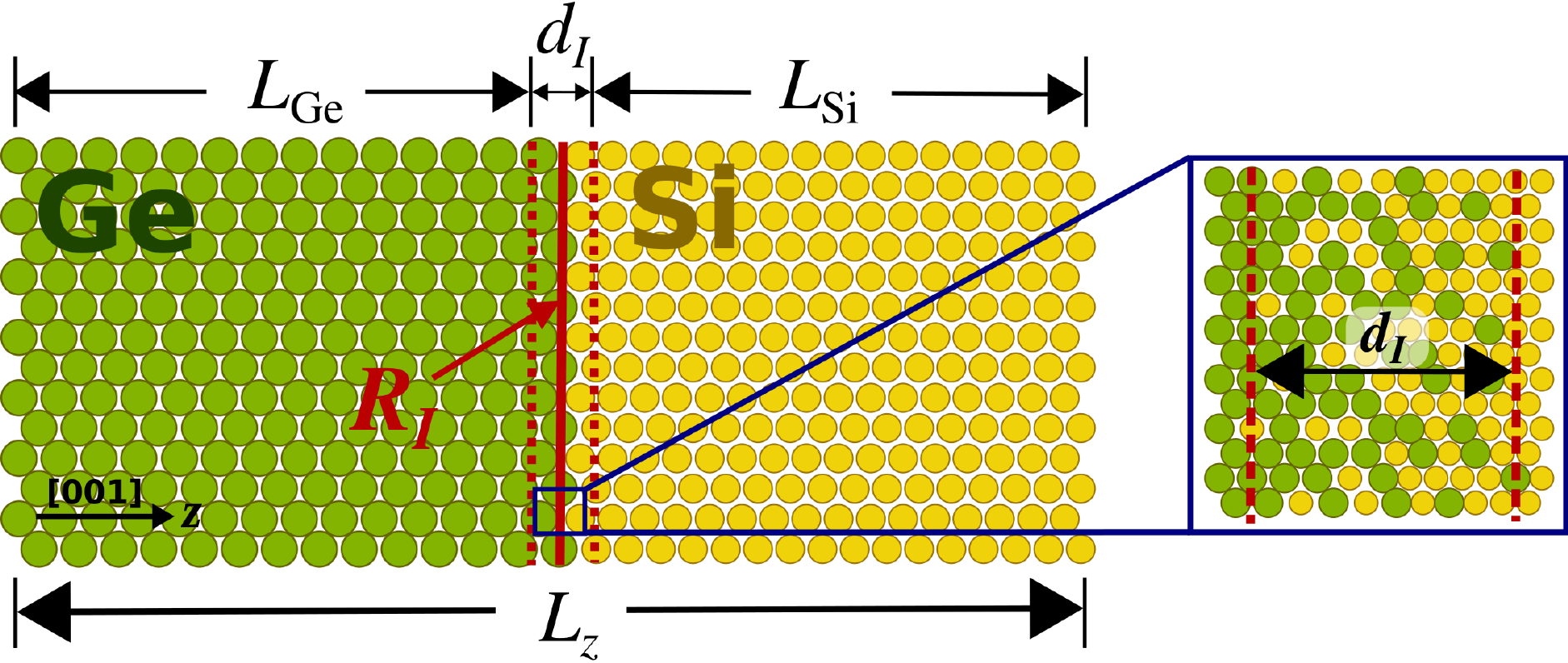}
       \caption{Schematic representation of a Ge/Si interface (in the (001) crystallographic plane) with boundary thickness $d_I$. The lattice spacing in in-plane direction (orthogonal to thermal transport) of Si has been adopted from the equilibrium lattice spacing of Ge ($a_{\mathrm{Ge},0}$=5.6567 \AA). The out-of-plane lattice spacing of Si has been calculated from the elastic constants ($a_\mathrm{Si,\perp}$=5.1913 \AA).}
       \label{fig:cell}
\end{figure}

\paragraph*{Creation of Si/Ge interfaces}
A schematic image of a Si/Ge simulation cell is represented in Fig. \ref{fig:cell}. The Si/Ge crystals are oriented with the crystallographic (001) plane orthogonal to the heat flux (in-plane direction). To account for periodic boundary conditions in the in-plane direction, the in-plane lattice spacing has to be equal for both Ge and Si sections. Motivated by experimental studies,\cite{FerreLlin2013} where SiGe alloys have been grown on pure Ge, we have imitated growth of Si on crystalline Ge where the equilibrium lattice parameter of Ge ($a_{\mathrm{Ge},0}$) has been adopted for the in-plane lattice spacing of Si. The out-of-plane lattice parameter of Si ($a_{\mathrm{Si},\perp}$) has been determined from the elastic properties of Si according to

\begin{equation}
\label{eq:a_off}
a^{\rm Si}_\perp=a^{\rm Si}\left[1-2\left(\dfrac{C_{12}}{C_{11}}\right)^{\rm Si}\left(\dfrac{a^{\rm Ge,0}}{a^{\rm Si,0}}-1\right)\right]
\end{equation}

with $C_{11}$ and $C_{12}$ of 142.54 and 75.38 GPa, respectively, resulting in $a_{\mathrm{Si},\perp}$=5.1913 \AA. The elastic constants $C_{11}$ and $C_{12}$ applying for the Tersoff potential have been determined from the second derivatives of the total energy with respect to deformation.
Pseudomorphic Si samples with these properties will be referred to as p-Si in the following.

In samples with a finite interface thickness $d_I$, several atomic layers of a Si$_x$Ge$_{1-x}$ alloy have been added at any interface. The Si concentration $x$ in the finite interfaces has been gradually increased every two atomic layers from the pristine Ge section to the pure p-Si section. The interfacial spacing has been determined equivalent to p-Si based on eq. \eqref{eq:a_off} and has been adapted for each Si concentration $x$. An interfacial spacing of 0.5, 1 and 2 nm has been simulated, corresponding to a total of four, eight and 16 atomic layers, respectively.

\subsection{Determination of thermal boundary resistance}
\label{sec:tbr_theory}
The overall thermal conductivity $\kappa_\mathrm{all}$ of a heterogeneous system, such as the heterostructure shown in Fig. \ref{fig:cell}, respectively its thermal resistance ($R_\mathrm{all}=\frac{L_z}{\kappa_\mathrm{all}}$), can be described as a connection of series of resistances. For the systems calculated here (Fig. \ref{fig:cell}) this can be expressed according to 
\begin{equation}
\label{eq:r_all}
R_\mathrm{all}=\frac{L_z}{\kappa_{\mathrm{all}}}=\frac{L_{\rm Ge}}{\kappa_{\rm Ge}\left(L_\mathrm{Ge}\right)}+\frac{L_{\rm Si}}{\kappa_{\mathrm{\text{p-}Si}}\left(L_\mathrm{\text{p-}Si}\right)}+2R_I
\end{equation}

where $L_z$, $L_{\rm Ge}$ and $L_{\rm Si}$ are the total simulation cell length and the length of the crystalline Ge and the pseudomorphic Si part, respectively (see Fig. \ref{fig:cell}). $\kappa_\mathrm{Ge}$ and $\kappa_\mathrm{\text{p-}Si}$ represent the thermal conductivities of crystalline Ge and p-Si at their cell length $L_\mathrm{Ge}$ and $L_\mathrm{Si}$, respectively. The thermal boundary resistance $R_I$ enters in the equation twice due to the periodic boundary conditions of the simulation cell.

\paragraph*{Thermal conductivity of $\kappa\left(L\right)$ of Ge and Si}
Following eq. \eqref{eq:r_all}, the length-dependent thermal conductivities of the pure systems $\kappa_\mathrm{Ge}$ and $\kappa_\mathrm{\text{p-}Si}$ have to be known for the determination of $R_\mathrm{all}$. This has been done using the standard AEMD approach for a homogeneous system\cite{Lampin2013,Melis2014a,Hahn2014} as described above. Therefore, the sample length of these systems has been varied from 100 nm ($\sim$200$a_0$ ) to $\sim$1 $\upmu$m (2000$a_0$). The behavior of 1/$\kappa$ to 1/$L_z$ has been approximated by a linear function $\frac{1}{\kappa}=\frac{1}{\kappa_\infty}\left(1+\frac{\lambda}{L_z}\right)$, where $\kappa_\infty$ is the bulk thermal conductivity and $\lambda$ can be defined as characteristic length of the phonon transport. The linear approximation is a common way to describe the dependency of $\kappa$ on the sample length and is approved for systems where phonon properties are approximated well by an average value.\cite{Schelling2002,Sellan2010} It has been shown to give reasonable estimations of the bulk thermal conductivity in Si/Ge systems.\cite{Schelling2002,Landry2009,Hahn2014}

\begin{figure}[tb]
\centering
  \includegraphics[angle=270, width=0.45\textwidth]{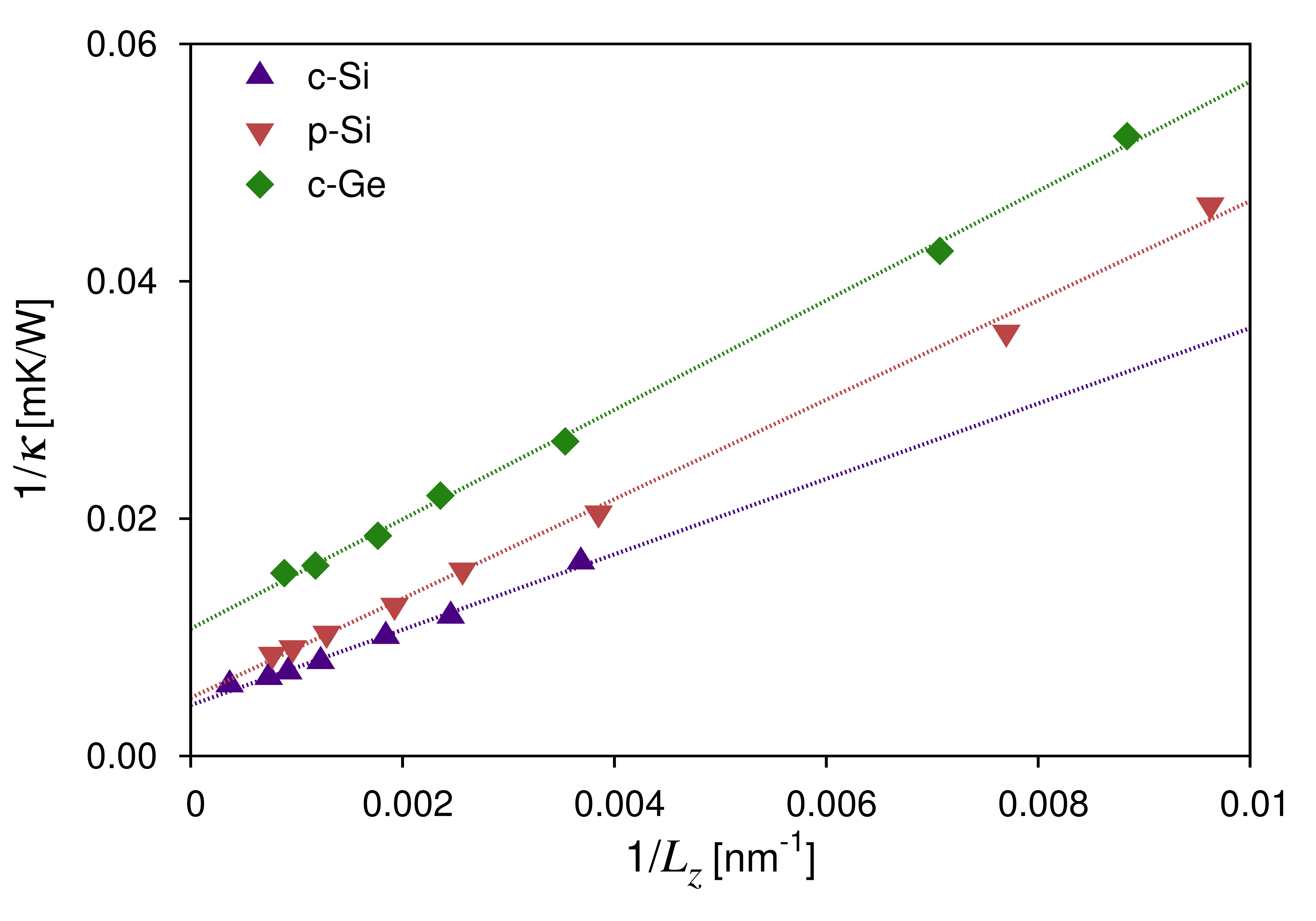}
       \caption{Inverse of the thermal conductivity 1/$\kappa$ as a function of the inverse sample length 1/$L_z$ of crystalline Ge (c-Ge, diamonds), crystalline Si (c-Si, triangles up) and pseudomorphic Si (p-Si, triangles down). The bulk thermal conductivity $\kappa_\infty$ of these materials has been approximated based on a linear relationship between 1/$\kappa$ and 1/$L_z$. It resulted in 93.3, 233.4 and 204.3 W/mK for c-Ge, c-Si and p-Si, respectively.}
       \label{fig:K_bulk}
\end{figure}

The length-dependent thermal conductivity of crystalline Ge, Si and pseudomorphic Si is shown in Fig. \ref{fig:K_bulk} with bulk thermal conductivities estimated to be 93.3, 233.4 and 204.3 W/mK, respectively (Table \ref{tab:K_vs_L}). The strain applied on the pseudomorphic Si as a result of the non-equilibrium in-plane lattice spacing thus reduces the thermal conductivity of the crystalline Si by 12.4 \%.

\begin{table}[bt]
\caption{Optimized parameters $\kappa_\infty$ and $\lambda$, describing the length-dependent thermal conductivity ($\frac{1}{\kappa}=\frac{1}{\kappa_\infty}\left(1+\frac{\lambda}{L_z}\right)$) of crystalline Ge, Si and pseudomorphic Si.}
\label{tab:K_vs_L}
\begin{center}
\begin{tabular}{ccc}
\hline \hline\noalign{\smallskip}
& $\kappa_\infty$ [$\frac{\mathrm{W}}{\mathrm{mK}}$] & $\lambda$ [nm] \\
\noalign{\smallskip}
\hline \noalign{\smallskip}
c-Gi & 93.3 $\pm$ 3.5 & 430 $\pm$ 24.4 \\
c-Si & 233.4 $\pm$ 13.4 & 742 $\pm$ 72.4 \\
p-Si & 204.3 $\pm$ 20.6 & 855 $\pm$ 108\\
\noalign{\smallskip}
\hline \hline
\end{tabular}
\end{center}
\end{table}

\paragraph*{Overall thermal conductivity}
The overall thermal conductivity is a summation of several phonon transport effects (eq. \eqref{eq:r_all}). Accordingly, the assumption of average phonon properties as in the case of the pure Ge and Si materials is unfounded. To account for the non-linear effects in $\kappa_\mathrm{all}$, the Taylor expansion of 1/$\kappa$ to 1/$L_z$ has been extended to the second order term.

\begin{equation}
\label{eq:kall}
\frac{1}{\kappa_\mathrm{all}}=\frac{1}{\kappa_{\mathrm{all},\infty}}\left(1+\frac{\lambda_\mathrm{all}}{L_z}+\frac{\mu_\mathrm{all}}{L_z^2}\right)
\end{equation}

\begin{figure}[tb]
\centering
  \includegraphics[width=0.45\textwidth]{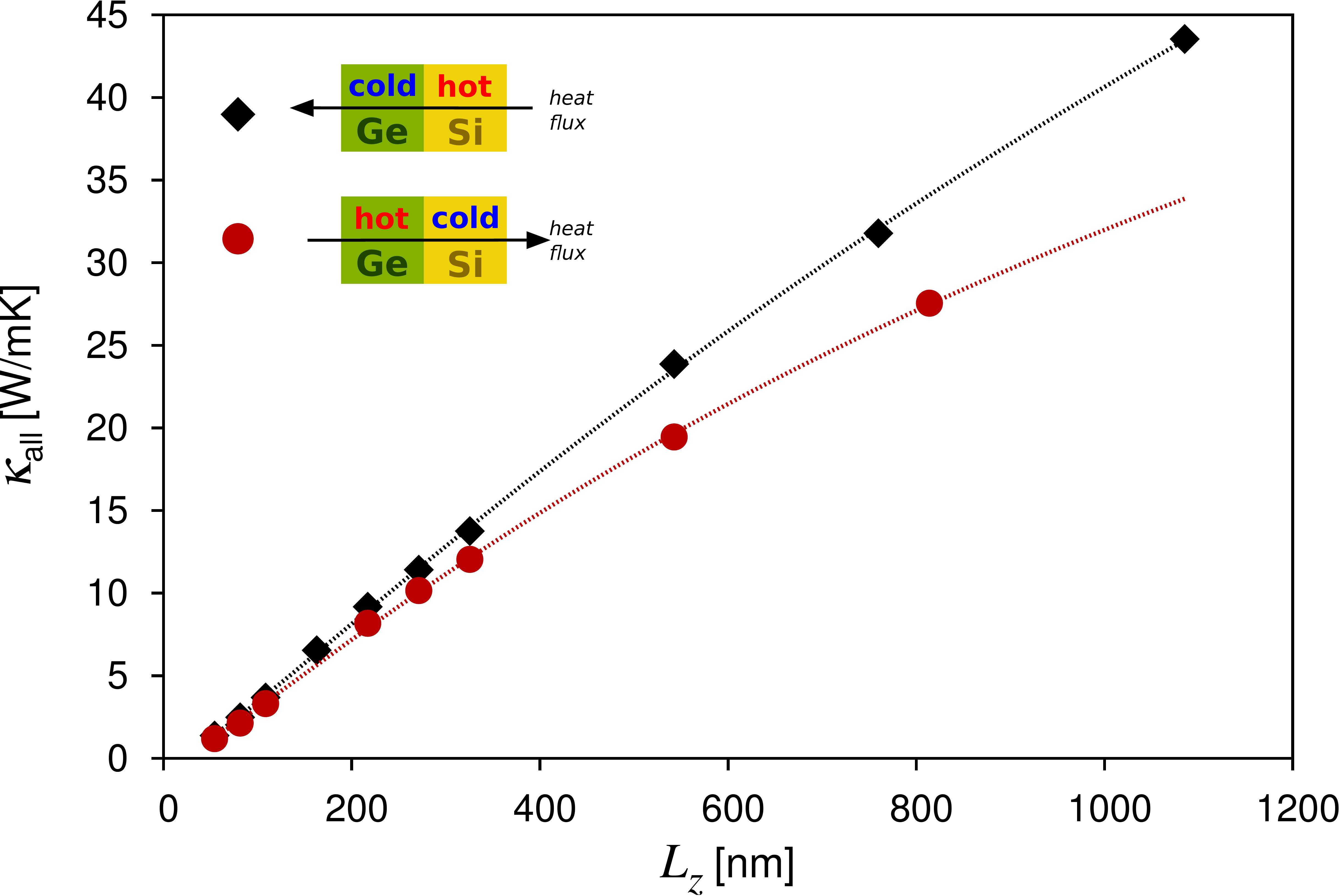}
       \caption{Total thermal conductivity $\kappa_\mathrm{all}$ as a function of the sample size $L_z$ of Si/Ge interfaces with a sharp interface. Thermal transport has been simulated from Si to Ge (diamonds) and vice versa (circles). The behavior of $\kappa$ to $L_z$ has been approximated with a 2$^{\mathrm{nd}}$ order polynomial function (dashed line).}
       \label{fig:Ktot_L}
\end{figure}

In fact, a linear approximation results in a negative bulk thermal conductivity $\kappa_{\mathrm{all},\infty}$, whereas the fitted second order function approximates the calculated values of $\kappa_\mathrm{all}$ perfectly as shown in Fig. \ref{fig:Ktot_L} for sharp interfaces.

\paragraph*{TBR estimation}
In order to estimate the thermal boundary resistance the effects of a finite simulation cell have to be eliminated. Therefore, eq. \eqref{eq:r_all} is rewritten as 

\begin{equation}
\label{eq:r_all_2}
\frac{1}{\kappa_{\mathrm{all}}}=\frac{\alpha_{\rm Ge}}{\kappa_{\rm Ge}\left(L_\mathrm{Ge}\right)}+\frac{\alpha_{\rm Si}}{\kappa_{\mathrm{\text{p-}Si}}\left(L_\mathrm{\text{p-}Si}\right)}+2\frac{R_I}{L_z}
\end{equation}

where the length of the Ge and Si part is expressed with respect to the total cell length as $\alpha_\mathrm{Ge}=\frac{L_\mathrm{Ge}}{L_z}$ and $\alpha_\mathrm{Si}=\frac{L_\mathrm{Si}}{L_z}$, respectively.

At infinite sample size ($L_z\rightarrow\infty$), the length-dependent $\kappa_\mathrm{all}$, $\kappa_\mathrm{Ge}$ and $\kappa_\mathrm{\text{p-}Si}$ converge to their bulk thermal conductivities $\kappa_{\mathrm{all},\infty}$, $\kappa_{\mathrm{Ge},\infty}$ and $\kappa_{\mathrm{\text{p-}Si},\infty}$, respectively, and the last term $\frac{R_I}{L_z}$ of eq. \eqref{eq:r_all_2} vanishes. With these assumptions, the bulk thermal conductivities are related according to eq. \eqref{eq:k_inf}.
 
\begin{equation}
\frac{1}{\kappa_{\mathrm{all},\infty}}=\frac{\alpha_{\mathrm{Ge}}}{\kappa_{\mathrm{Ge},\infty}} + \frac{\alpha_{\mathrm{Si}}}{\kappa_{\mathrm{\text{p-}Si,\infty}}}
\label{eq:k_inf}
%\frac{1}{\kappa_{\mathrm{all},\infty}}=\frac{\alpha_{\mathrm{Ge}}}{\kappa_{\mathrm{Ge},\infty}} +\frac{\alpha_{\mathrm{Si}}}{\kappa_{\mathrm{\text{p-}Si,\infty}}}
\end{equation}

With the linear approximation of 1/$\kappa\left(1/L\right)$ in c-Ge and p-Si and the quadratic behavior of 1/$\kappa_\mathrm{all}\left(1/L\right)$ for the overall thermal conductivity (eq. \eqref{eq:kall}), \eqref{eq:r_all} can be rewritten as follows

\begin{equation}
\label{eq:r_all_3}
\begin{split}
R_I=\frac{1}{2}\left( C_1\times L_z + C_2 + C_3 \times \frac{1}{L_z}\right) \\
\text{where} \hspace{2ex}C_1=\frac{1}{\kappa_{\mathrm{all},\infty}}-\frac{\alpha_\mathrm{Ge}}{\kappa_{\mathrm{Ge},\infty}}-\frac{\alpha_\mathrm{Si}}{\kappa_{\mathrm{\text{p-}Si},\infty}}\\
C_2=\frac{\lambda_\mathrm{all}}{\kappa_{\mathrm{all},\infty}}-\frac{\lambda_\mathrm{Ge}}{\kappa_{\mathrm{Ge},\infty}}-\frac{\lambda_\mathrm{\text{p-}Si}}{\kappa_{\mathrm{\text{p-}Si},\infty}}\\
C_3=\frac{\mu_\mathrm{all}}{\kappa_{\mathrm{all},\infty}}.
\end{split}
\end{equation}

As a result, the thermal boundary resistance this simplifies to

\begin{equation}
\label{eq:r_inf}
R_I=\frac{1}{2}\left(C_2+C_3\times\frac{1}{L_z}\right)=R_{I,\infty}\left(1+\frac{\lambda_I}{L_z}\right)
\end{equation}

where $R_{I,\infty}$ is the converged TBR at bulk conditions ($L_z\rightarrow\infty$) and $\lambda_I$ is the characteristic phonon length of the interface.
To determine the parameters of $R_I\left(L_z\right)$, $R_I$ has been calculated at each data point from the overall thermal conductivity $\kappa_\mathrm{all}$ (eq. \eqref{eq:kall}) and previously determined length-dependent $\kappa$ of the pristine Ge and p-Si materials. The obtained values have then been fitted to eq. \eqref{eq:r_inf}.

%The error in $R_I$ has been estimated according to the error formula of propagation \cite{Ku1966}.
%\clearpage

\section{Results and Discussion}
\subsection{TBR at sharp interface}
\label{sec:tbr_sharp}
\begin{figure}[tb]
\centering
%\resizebox{0.45\textwidth}{!}{\includegraphics[angle=270]{R_L_Si2Ge}}
  \includegraphics[angle=270, width=0.45\textwidth]{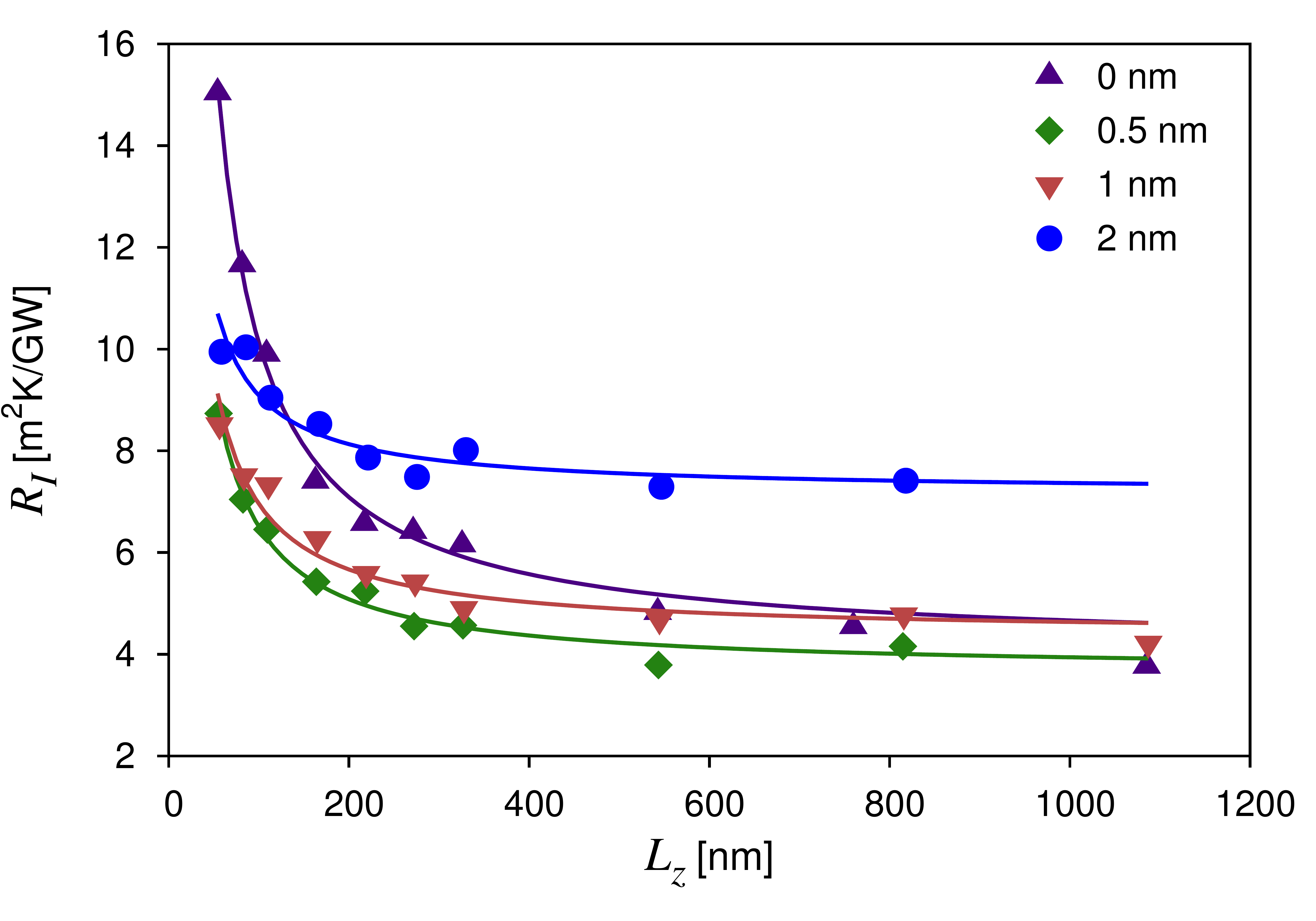}
       \caption{Thermal boundary resistance $R_I$ as a function of the sample size $L_z$. The behavior of $R_I$ to $L_z$ follows a linear reciprocal trend $R_I=R_{I,\infty}\left(1+\frac{\lambda _I}{L_z}\right)$.}
       \label{fig:R_L}
\end{figure}

In Fig. \ref{fig:R_L}, the calculated AEMD values of $R_I$ as a function of the simulation cell length $L_z$ together with their fits by eq. \ref{eq:r_inf} are shown for Si/Ge interfaces with various boundary thickness $d_I$ where thermal transport has been simulated from Si to Ge.

The bulk TBR $R_{I,\infty}$ of a sharp interface has been calculated to be 3.76$\times 10^{-9}$ m$^2$K/W (Table \ref{tab:R_inf}). 
%In a previous study,\cite{Landry2009a} the TBR has been estimated from Laundauer-like expressions for phonon transport assuming elastic phonon scattering and from non-equilibrium molecular dynamics simulations applying the direct method and calculating the thermal resistance from the temperature gradient across the interface. 
Several theoretical methods have been used in a previous study for the estimation of the TBR.\cite{Landry2009a} Among others, non-equilibrium molecular dynamics simulations have been used where the thermal resistance is calculated from the temperature gradient across the interface. Using this approach, the thermal boundary resistance at 300 K resulted in 3.1$\times 10^{-9}$ m$^2$K/W.
Despite the differences in the applied methods in ours and the previous work (which include a different lattice spacing, a different pair potential and a different approach to calculate $R_I$), this is in very good agreement with our results. Furthermore, the value obtained from both MD-based methods are in reasonable agreement with the TBR calculated from the theoretical, Laundauer-like expressions for phonon scattering (3.0$\times 10^{-9}$ m$^2$K/W) and the results using the DMM (2.4$\times 10^{-9}$ m$^2$K/W)\cite{Landry2009a} where approximations of the nature of phonon scattering (such as elastic scattering) have to be applied. The AEMD method used here thus provides a robust approach to reliably estimate the TBR without the necessity of approximations for phonon scattering and verifies the applicability of a mean thermal conductivity which is assumed to be uniform in the direction of the heat transport.

When the heat flow is inverted, i.e. the Ge region is initially equilibrated at $T_h$=400 K and p-Si at $T_c$=200 K, the thermal boundary resistance increases to 5.78$\times 10^{-9}$ m$^2$K/W (Table \ref{tab:R_inf}).
Similar results have been shown previously for the heat transport in SiGe nanowires.\cite{Rurali2014} Using non-equilibrium MD simulations, the heat transport has been shown to be higher for a heat flow direction from Si to Ge than vice versa.

The increase by $\sim$ 50 \% can be explained by the different thermal conductivities of the pristine Ge and p-Si materials. Silicon has a higher thermal conductivity than Ge. Accordingly, the inflow of heat from the p-Si region to the interface is facilitated over its outflow through the Ge region which leads to a higher heat pressure at the Si/Ge interface. A higher heat pressure in this sense is comparable to a higher temperature of the system. This is comparable to previous results where the TBR has been shown to decrease with increasing process temperature from 300 to 1000 K.\cite{Landry2009}

\subsection{Effect of interface thickness on TBR}
\label{sec:d_effect}
The morphology of the Si/Ge interface has been varied by changing its thickness $d_I$ in the range 0 nm$\leq d_I \leq$ 2 nm. At a finite thickness, the interface has been constructed of a Si$_x$Ge$_{1-x}$ alloy in which the Si concentration $x$ has been increased gradually from the Ge region to the Si region as described in section \ref{sec:methods}.

\begin{figure}[tb]
\centering
  \includegraphics[width=0.45\textwidth]{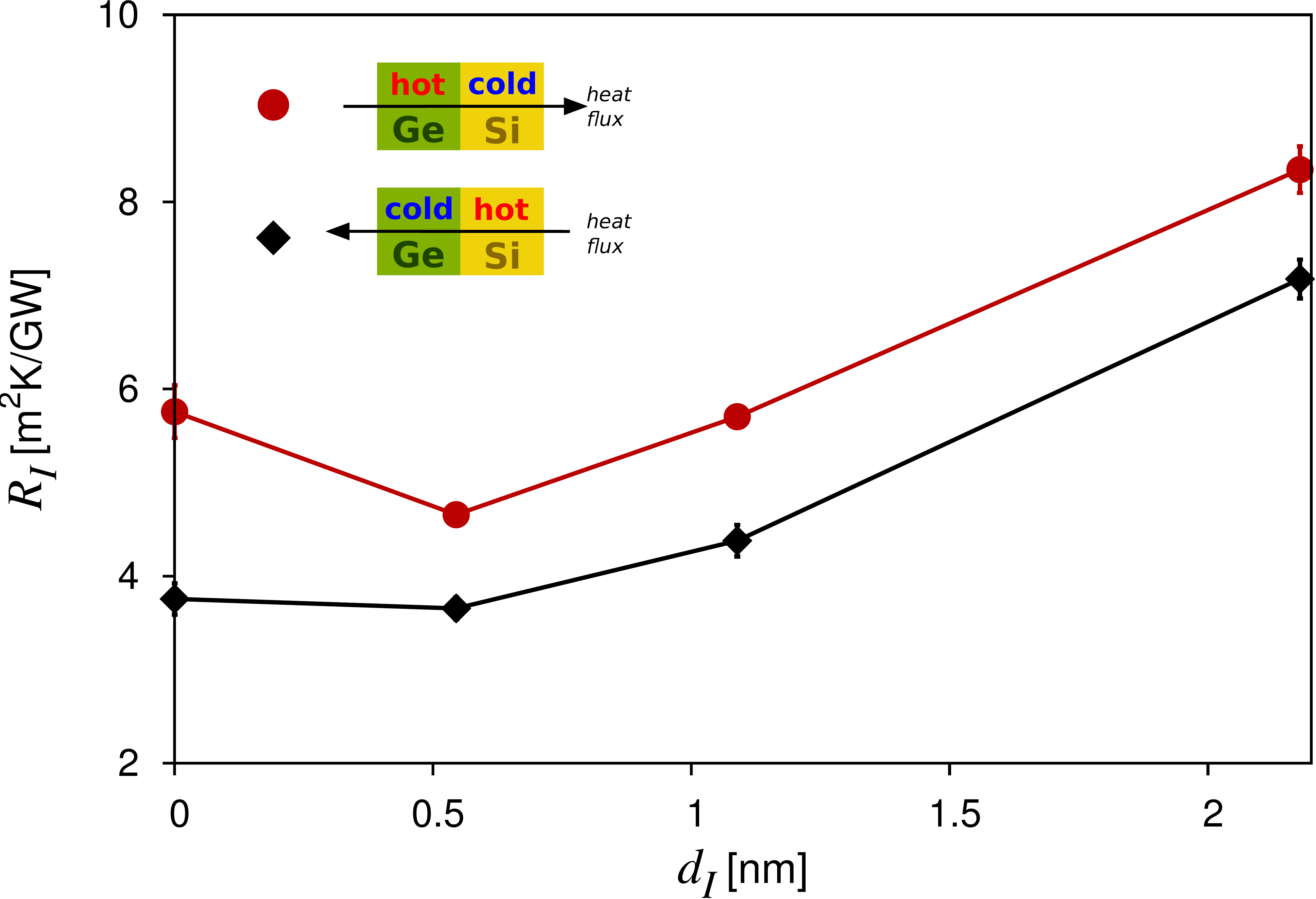}
       \caption{Bulk thermal boundary resistance $R_{I,\infty}$ as a function of the interfacial thickness $d_I$ for thermal transport from Si to Ge (diamonds) and from Ge to Si (circles).}
       \label{fig:R_d}
\end{figure}

Switching from a vanishingly thin interface ($d_I=0$ nm) to a finite thickness of 0.5 nm results in a slight decrease ($\Updelta R_I=0.1\times10^{-9}$ m$^2$K/W, see Fig. \ref{fig:R_d} and Table \ref{tab:R_inf}) of the TBR in the case of thermal transport from p-Si to Ge.
This effect is more pronounced when the heat flow is inverted ($\Updelta R_I=1.1\times10^{-9}$ m$^2$K/W). The introduction of an interface with finite thickness reduces the lattice and mass mismatch between the pure materials and the SiGe alloy. As a result, the TBR is reduced.

Further boundary growth, however, results in a steady increase of the thermal boundary resistance. For thermal transport from p-Si to Ge, it raises up to 7.18 $\times10^{-9}$ m$^2$K/W at a boundary thickness of 2 nm (Table \ref{tab:R_inf}). This is in agreement with previous results of the TBR in SiGe nanowires calculated by non-equilibrium MD simulations showing TBR to increase when the Si/Ge interfaces thickness increases from abrupt to 5 and 15 nm.\cite{Rurali2014}
When the heat flow direction is inverted, it reaches 8.37 $\times10^{-9}$ m$^2$K/W. The increase in TBR with increasing thickness can be explained by increased phonon scattering inside the boundary layer. For a sufficiently large boundary, this effect dominates the reduced lattice and mass mismatch and the TBR exceeds the value of the sharp interface. In the case of thermal transport from Si to Ge this is already the case at $d_I$=1 nm while for heat flow from Ge to Si, it is observed at $d_I=$2 nm.

The characteristic phonon length $\lambda_I$ (eq. \ref{eq:r_inf}), on the other hand, is hardly affected from the heat flow direction. Except for the sharp interface, $\lambda_I\left(d\right)$ is within 10\% for both directions of the heat flow (Table \ref{tab:R_inf}). It decreases monotonously from 78.1 (76.3) to 26.2 (27.8) nm with increasing boundary thickness from 0.5 to 2 nm for a heat flow from p-Si to Ge (Ge to p-Si). The characteristic length does not directly correspond to any phonon property, however, it gives an indication. Here, its decrease with increasing boundary thickness can be regarded as a decrease in the phonon mean free path, presumably resulting from enhanced phonon scattering inside the boundary layer.

\begin{table}[bt]
\caption{Optimized parameters $R_{I,\infty}$ and $\lambda_I$ according to eq. \ref{eq:r_inf} for thermal transport from Si to Ge and vice versa for Si/Ge interfaces with different boundary thickness $d_I$.}
\label{tab:R_inf}
\begin{center}
\begin{tabular}{ccc}
\hline \hline\noalign{\smallskip}
\multicolumn{1}{c}{}& \multicolumn{2}{c}{Si $\rightarrow$ Ge} \\
$d_I$ [nm] & $R_{I,\infty}$ [$\frac{\mathrm{m}^2\mathrm{K}}{\mathrm{GW}}$] & $\lambda_I$ [nm]\\
\noalign{\smallskip}
\hline\noalign{\smallskip}
%\rule{0pt}{3ex}
0	& 3.76 $\pm$ 0.17 & 167.2 $\pm$ 13.6 \\
0.5	& 3.66 $\pm$ 0.12 & 78.1 $\pm$ 6.4 \\
1	& 4.38 $\pm$ 0.17 & 58.9 $\pm$ 7.5 \\
2	& 7.18 $\pm$ 0.21 & 26.6 $\pm$ 4.5 \\
\noalign{\smallskip}
\hline\noalign{\smallskip}
\multicolumn{1}{c}{}& \multicolumn{2}{c}{Ge $\rightarrow$ Si} \\
$d_I$ [nm] & $R_{I,\infty}$ [$\frac{\mathrm{m}^2\mathrm{K}}{\mathrm{GW}}$] & $\lambda_I$ [nm]\\
\noalign{\smallskip}
\hline\noalign{\smallskip}
%\rule{0pt}{3ex}
0	&  5.76 $\pm$ 0.28 & 115.6 $\pm$ 11.8\\
0.5	&  4.66 $\pm$ 0.12 & 76.3 $\pm$ 5.0\\
1	&  5.70 $\pm$ 0.08 & 53.7 $\pm$ 2.4\\
2	&  8.34 $\pm$ 0.25 & 27.8 $\pm$ 4.7\\
\noalign{\smallskip}
\hline \hline
\end{tabular}
\end{center}
\end{table}

\subsection{Rectification}
\label{sec:rect}
\begin{figure}[tb]
\centering
  \includegraphics[angle=270, width=0.45\textwidth]{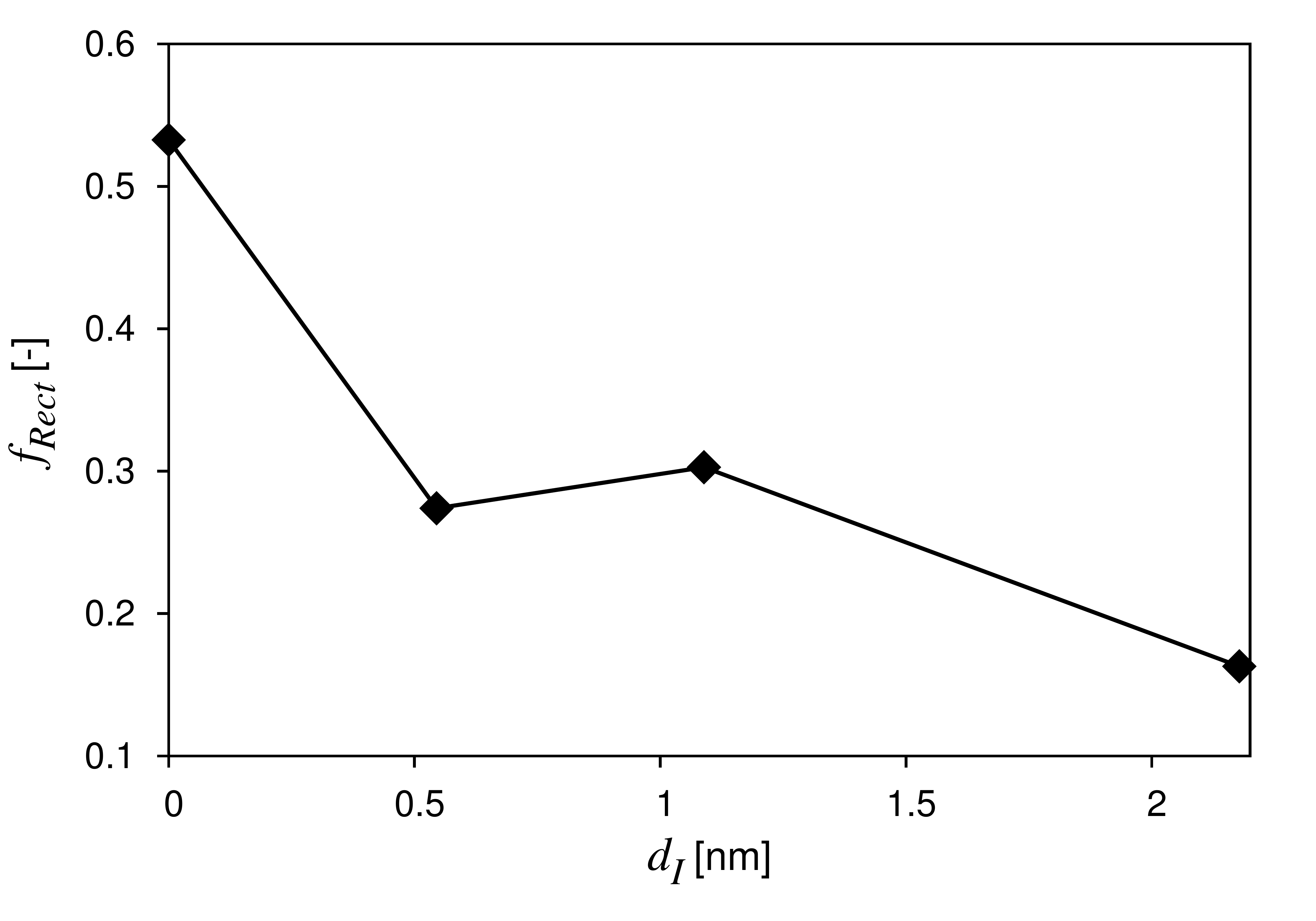}
       \caption{Rectification $f_{Rect}$ of the bulk thermal boundary resistance of Si/Ge interfaces as a function of the interfacial thickness $d_I$. It has been calculated according to eq. \ref{eq:rect}.}
       \label{fig:Rect}
\end{figure}

Thermal rectification can be understood in analogy to an electrical diode. A material is defined as a thermal rectifier if the magnitude of the heat flow is different when inverting the heat flow direction.\cite{Dames2009} Several expressions have been used recently to specify the thermal rectification. Most commonly it is defined as the difference of the magnitude of backward and forward heat flow divided by the smaller one.\cite{Rurali2014,Dames2009, Pereira2013a, Wang2014} 
Within the AEMD framework, the magnitude of the heat flow is not directly calculated. However, assuming temperature independent thermal conductivities of the pure Ge and Si materials, justified here by a narrow $\Delta T$, the heat flow in one or the other direction depends only on the thermal boundary resistance (eq. \ref{eq:hc}). Following this, we define the rectification factor $f_{Rect}$ of the TBR as the difference between the thermal resistance of backward (Ge to Si, $R_b$) and forward (Si to Ge, $R_f$) heat flow divided by the one of forward heat flow (eq. \ref{eq:rect}).

\begin{equation} 
f_{Rect}=\frac{R_b-R_f}{R_f} \label{eq:rect}
\end{equation}
Thermal rectification $f_{Rect}$ as calculated here includes only the contribution of the TBR to the heat flow but neglects the effect of temperature dependent heat transport in the crystalline segments.
The rectification of the TBR remarkably drops from 0.53 to 0.27 when the sharp boundary between Si and Ge is smoothed to a boundary layer with finite thickness of 0.5 nm (Fig. \ref{fig:Rect}). Further increase of the boundary thickness only had little effect on the thermal rectification. This is in agreement with what has been shown previously for thermal rectification of a two-segment device. Rectification in such a device is increased when the asymmetry and mismatch of the different materials in the two segments is more pronounced. The introduction of a boundary layer  with finite thickness consisting of a Si$_x$Ge$_{1-x}$-alloy reduces the asymmetry and the mismatch between the materials that are directly connected, thus resulting in a decrease of the rectification factor.

%\clearpage

\section{Conclusions}
Approach-to-equilibrium molecular dynamics (AEMD) simulations have been applied to determine the thermal boundary resistance (TBR) of Si/Ge interfaces. The overall thermal conductivity has been described by a connection of series of several resistances, including the resistance of the interface. Its dependence on the sample length has been described by a reciprocal quadratic behavior to account for non-linear effects. With this assumption the length dependent TBR could be expressed by a reciprocal linear function which converges to a bulk TBR for infinitely large simulation cells.
The bulk TBR for thermal transport from Si to Ge at 300 K resulted in 3.76$\cdot10^{-9}$ m$^2$K/W in agreement with previous calculations using non-equilibrium molecular dynamics simulations.

The effect of the interface morphology on the bulk TBR has been evaluated focusing on the variation of the interfacial thickness. The interface, consisting of a Si$_x$Ge$_{1-x}$ alloy, has been increased from a sharp interface to one with a thickness of 2 nm. TBR is found to slightly decrease when switching from a sharp interface to an interface with finite thickness. This can be explained by reduced mismatch and asymmetry between the pure Si and Ge and the interface consisting of a SiGe alloy. Further increase of the interface thickness, however, leads to enhanced phonon scattering and results again in an increase of the TBR.

Furthermore, the effect of heat flux inversion has been investigated simulating thermal transport from Ge to Si. In this case, thermal boundary resistance has been found to be higher than for thermal transport from Si to Ge, independent of the thermal boundary thickness. From these results, the rectification of the Si/Ge interface has been determined. It is most pronounced when the interface is infinitely sharp (0.53) and decreases significantly when the interface is smoothed over a finite thickness.

These results give insight into the thermal transport properties of Si/Ge interfaces, indicating a notable influence of the composition and morphology of the interface on the thermal boundary resistance. At a certain boundary thickness phonon scattering dominates over the effects of reduced mass mismatch at the interface, leading to an increase of the thermal boundary resistance which even exceeds the value at a sharp interface. It is thus suggested that reduction of thermal conductivity is more effective at extended interfaces with a certain thickness.

\begin{acknowledgments}
This work is financially supported by the SNF grant with the project number P2ZHP2\_148667. Simulations have been conducted on the HPC resources of CINECA under the project ISCRA\_THETRASI.
\end{acknowledgments}

\bibliographystyle{./apsrev} % Title is link if provided

%
%\bibliographystyle{/home/konstanze/lib/template/latex/backmatter/krh_2} % Title is link if provided
%%\renewcommand{\refname}{References} % changes the header; default: Bibliography
%
%%\bibliography{../10_backmatter/Thesis} % adjust this to fit your BibTex file
%\bibliography{/home/konstanze/Literature/Bibtex/ThermoElectric} % adjust this to fit your BibTex file
%\bibliography{short,ThermoElectric_2_AltJ}
%\end{small}
%

\end{document}